\documentclass[aps,pra,twocolumn,superscriptaddress,10pt]{revtex4-2}
\usepackage[utf8]{inputenc}
\usepackage{multirow,amsthm,amssymb,amsbsy,amsmath,epsfig,epstopdf,bm,float}
\usepackage{bm}
\usepackage{graphicx}
\usepackage{booktabs}
\usepackage{verbatim}
\usepackage{dcolumn}
\usepackage{color}
\usepackage[dvipsnames,table,xcdraw]{xcolor}
\usepackage[normalem]{ulem}
\usepackage{soul}
\usepackage{braket}
\usepackage{subfig}
\usepackage[export]{adjustbox}
\usepackage{color}

\usepackage{numprint} 

\begin{document}
\title{Quantum fidelity kernel with a trapped-ion simulation platform}
\author{Rodrigo Mart\'inez-Pe\~na}
\email{rodrigo.martinez@dipc.org}
\affiliation{Instituto de F\'{i}sica Interdisciplinar y Sistemas Complejos (IFISC, UIB-CSIC), Campus Universitat de les Illes Balears E-07122, Palma de Mallorca, Spain}
\affiliation{Donostia International Physics Center, Paseo Manuel de Lardizabal 4, E-20018 San Sebastián, Spain}

\author{Miguel C. Soriano}
\affiliation{Instituto de F\'{i}sica Interdisciplinar y Sistemas Complejos (IFISC, UIB-CSIC), Campus Universitat de les Illes Balears E-07122, Palma de Mallorca, Spain}

\author{Roberta Zambrini}
\affiliation{Instituto de F\'{i}sica Interdisciplinar y Sistemas Complejos (IFISC, UIB-CSIC), Campus Universitat de les Illes Balears E-07122, Palma de Mallorca, Spain}

\date{ \today }

\begin{abstract}

Quantum kernel methods leverage a kernel function computed by embedding input information into the Hilbert space of a quantum system. However, large Hilbert spaces can hinder generalization capability, and the scalability of quantum kernels becomes an issue. To overcome these challenges, various strategies under the concept of inductive bias have been proposed. Bandwidth optimization is a promising approach that can be implemented using quantum simulation platforms. We propose trapped-ion simulation platforms as a means to compute quantum kernels, filling a gap in the previous literature and demonstrating their effectiveness for binary classification tasks. We compare the performance of the proposed method with an optimized classical kernel and evaluate the robustness of the quantum kernel against noise. The results show that ion trap platforms are well-suited for quantum kernel computation and can achieve high accuracy with only a few qubits.

\end{abstract}


\maketitle
\section{Introduction}

The availability of noisy intermediate-scale quantum (NISQ) devices \cite{preskill2018quantum} has catalyzed a surge in several research areas such as machine learning, many-body physics, chemistry, and combinatorial optimization \cite{bharti2022noisy}.  In particular, quantum machine learning (QML) is a promising route toward quantum advantage, where some theoretical studies \cite{huang2021information,huang2021power,liu2021rigorous,sweke2021quantum} and experimental evidence \cite{huang2022quantum, cerezo2022challenges,melnikov2023quantum} have already indicated positive progress in this direction. While circuit-based NISQ computing implementations are developed worldwide \cite{bharti2022noisy}, an alternative approach exploits NISQ analog or hybrid designs, with examples ranging from quantum annealing \cite{das2008colloquium}, boson sampling \cite{brod2019photonic}, and QML itself \cite{markovic2020quantum,mujal2021opportunities}. In this work we will focus on the QML implementation of a kernel method in an analog simulation platform.
 
 Kernel methods refer to classification or regression algorithms that leverage a kernel function, denoted as $K(\bm{x},\bm{y})$, with $\bm{x},\bm{y}\in \mathcal{X}$, being $\mathcal{X}$ the input space. The kernel function transforms data points residing in the input space into a higher-dimensional space, called feature space.  This transformation usually facilitates the distinction between different classes of data when they are not linearly separable. A quantum kernel method is generally understood as the case where a kernel function is computed by embedding the input information into the Hilbert space of a quantum system (the feature space), offloading the optimization process for the ML algorithm to a classical computer \cite{schuld2019quantum}. The most common approach to computing quantum kernels is based on the quantum state fidelity, i.e., the inner product of input-dependent quantum states. Quantum advantage has been claimed as possible in realistic problems if the embedding of inputs into these quantum states is classically intractable \cite{schuld2019quantum,havlivcek2019supervised}. 

However, finding a kernel function that could provide a quantum advantage remains an open problem. One plausible strategy that could come to our minds is to propose a very large quantum Hilbert space as the feature space since, in principle, it would be much harder to simulate classically. But very large Hilbert spaces can hinder the generalization capability of quantum models \cite{kubler2021inductive}, performing similarly or even worse than classical ML models \cite{huang2021power}. Indeed, training a kernel-based model ensures the discovery of the optimal model parameters as the training landscape is convex. However, this assurance is based on the assumption that the quantum kernel can be efficiently obtained from quantum hardware. With an increasing number of qubits, the number of measurements required to evaluate the kernels to sufficiently high precision might scale exponentially, hindering trainability \cite{kubler2021inductive,thanasilp2022exponential,suzuki2022quantum}.

Several alternatives have been proposed to avoid this exponential scaling, all of them under the notion of what is called inductive bias \cite{kubler2021inductive,cerezo2022challenges}. An inductive bias is a restriction over the set of functions a given model can reproduce. They can be implemented in very diverse ways, such as projected quantum kernels \cite{huang2021power}, quantum Fisher kernels \cite{suzuki2022quantum}, exploiting structured-data encoding \cite{albrecht2023quantum}, and bandwidth optimization \cite{shaydulin2022importance,canatar2022bandwidth}.  While the former inductive biases can be problem-dependent, bandwidth optimization is a general strategy in which we simply tune the hyperparameters of our model. In particular, for classical data the decrease in performance with respect to the number of qubits can be compensated by the tuning of hyperparameters \cite{shaydulin2022importance,canatar2022bandwidth}, although it might be possible to find classical kernels that perform as well as them in some cases \cite{slattery2023numerical}. 

A promising choice for bandwidth optimization is quantum simulation platforms. Quantum simulations are believed to be a classically hard problem together with instantaneous quantum polynomial (IQP) random circuits or boson sampling \cite{harrow2017quantum}. Some analog simulation platforms have already been experimentally exploited for quantum kernels, such as Rydberg atoms \cite{albrecht2023quantum} and nuclear magnetic resonance (NMR) platforms \cite{kusumoto2021experimental}, while more platforms have been numerically studied, such as driven-dissipative spin chains \cite{bharti2022noisy}, a single Kerr nonlinear mode \cite{liu2023quantum} and quantum annealers \cite{yang2023analog}. Each physical platform has its own advantages and challenges, which not only depend on experimental details but also on the QML technique. 
For example, NMR experiments allow the expectation value of observables to be obtained with a single shot, thus removing the influence of the statistical noise on the experimental results. The need to deal with statistical noise can be a source of severe overhead in some QML techniques such as quantum reservoir computing \cite{mujal2021opportunities}. However, NMR platforms are difficult to initialise, whereas systems such as Rydberg atoms and trapped ions have very high fidelity \cite{morgado2021quantum,monroe2021programmable}. 

In this line of research, we propose to compute quantum kernels through trapped-ion simulation platforms, filling a gap in the previous literature. Trapped and laser-cooled atomic ions offer an excellent benchmark for simulating quantum spin models with interactions \cite{smith2016many,zhang2017observation,kaplan2020many}.  Optical fields can control the ions' Coulomb interaction, enabling customizable, long-range spin-spin interactions \cite{monroe2021programmable}. Ion trap quantum computers with logical quantum gates have already been used for classification tasks \cite{rudolph2022generation,johri2021nearest}, even with quantum kernel methods \cite{moradi2022error,suzuki2023quantum}.

In this work, we will introduce an inductive bias (i.e. a change in expressivity) in a transverse-field Ising model by applying bandwidth optimization, i.e., by tuning some of the model's hyperparameters. We show that this approach is suitable for solving binary classification tasks, comparing the performance with an optimized classical kernel. Our goal is to show via numerical simulations that an already available experimental platform can be harnessed for quantum kernel methods, demonstrating its robustness against noise. We introduce noise into the system in two different ways: decoherence with a depolarizing channel after the dynamics, and statistical noise with white noise at the entries of the kernel. For this, we require to use an error mitigation technique to keep the kernel as a positive semi-definite matrix, applying the shift method (see Sect. \ref{sect:QFK}). 

Our results show that ion trap platforms are perfectly suited for the computation of quantum kernels, being robust against different sources of noise. We evaluate the classification tasks for an increasing Hilbert space size, showing that only a few qubits are needed to obtain the best possible accuracy.

\section{Methods}

\subsection{Kernel methods and support vector machines}

In machine learning, kernel methods are algorithms used for classification or regression. They utilize a kernel function, denoted as $K(\bm{x},\bm{y})$, where $\bm{x},\bm{y}\in \mathcal{X}$ and $\mathcal{X}$ represents the input space, that transforms data points $\mathcal{X}$ to a higher-dimensional space, i.e. the feature space, usually facilitating the distinction between different classes of data. A function $K:\mathcal{X}\times \mathcal{X}\rightarrow \mathbb{R}$ is defined as a kernel function if it fulfills two conditions: 1) it is symmetric, such that $K(\bm{x},\bm{y})=K(\bm{y},\bm{x})$; and 2) it is a positive-semidefinite function. The latter is known as the Mercer condition \cite{mercer1909xvi}, which can be expressed as:
\begin{equation}
    \sum^M_{i=1}\sum^M_{j=1}K(\bm{x}_i,\bm{x}_j)c_ic_j\geq 0,
\end{equation}
where $M$ is the number of training samples and $\{c_i\}$ is the set of all possible real coefficients. A kernel function evaluated over pairs of data points defines a symmetric and positive semidefinite matrix whose elements are given by $K_{ij}:=K(\bm{x}_i,\bm{x}_j)$.

Here we will consider Support Vector Machines (SVM) as our kernel method \cite{vapnik1995nature}. Our starting point will be to introduce the simplest case: linear binary classification. This problem consists on drawing a straight line that separates two classes by assigning two different labels to them, like $y=\pm 1$, on opposite sites of the line. The decision function, i.e. the function that assigns a label for new data, can be determined as:
\begin{equation}
    D(\bm{x})=\bm{w}^\top\cdot \bm{x}+b,
\end{equation}
where $\bm{w}\in \mathbb{R}^m$ is a vector with the same dimension as the input and $b$ is an offset. The labeling is introduced by using the sign function:
\begin{equation}
    y(\bm{x})=\text{sgn}(\bm{w}^\top\cdot \bm{x}+b).
\end{equation}
Let us assume we have $M$ training inputs $\bm{x}_i$ belonging either to Class 1 with $y_i=1$ or Class 2 with $y_i=-1$. Since we assume the training data are linearly separable, the following inequality is fulfilled:
\begin{equation}
    y_iD(\bm{x})=y_i(\bm{w}^\top\cdot \bm{x}_i+b)\geq 1, \quad i=1,\dots,M.
\end{equation}
That is, the decision function yields $D(\bm{x}_i)\geq 1$ for $y_i=1$ and $D(\bm{x}_i)\leq -1$ for $y_i=-1$, separating both classes by the hyperplanes $D(\bm{x})=1$ and $D(\bm{x})=-1$. We can define a separating hyperplane between these two: $\bm{w}^\top\cdot \bm{x}+b=c$, for $-1<c<1$. The distance between the separating hyperplane and the training
data sample nearest to the hyperplane is called the margin. The generalization ability of this model depends on the location of the separating hyperplane, and the hyperplane with the maximum margin is denoted as the optimal one.  

In fact, $\bm{w}$ represents the orthogonal vector of the optimal separating hyperplane. If input data is bi-dimensional $\bm{w}$ defines a separation line, and in higher dimensions, it defines a separating hyperplane. However, a linear classifier will not suffice if the classification problem is nonlinear. Here comes the concept of the feature map $\phi( \bm{x})$, which embeds data points into a higher dimensional space (feature space) such that the modified inputs may become linearly separable. The new decision function becomes
\begin{equation} \label{eq:DF_w'}
    D(\bm{x})=\bm{w'}^\top\cdot \phi(\bm{x})+b,
\end{equation}
where $\bm{w'}$ is now the vector of the optimal separating hyperplane in the new feature space. The representer theorem \cite{scholkopf2002learning} states that the vector $\bm{w'}$ can be written as a sum over a subset $S$ of the training data points with real coefficients $\alpha_i$:
\begin{equation}
    \bm{w'}=\sum_{i\in S}\alpha_iy_i\phi(\bm{x}_i).
\end{equation}
In the case of SVM, the subset $S$ corresponds to the support vectors, that is, only the closest points to the decision boundary contribute. We write again the decision function with this new information,
\begin{equation}  
    D(\bm{x})=\sum_{i\in S}\alpha_iy_i\phi(\bm{x}_i)^\top\cdot \phi(\bm{x})+b,
\end{equation}
finally arriving at the introduction of the kernel, defined as an inner product between feature vectors: 
\begin{equation} \label{eq:K_phi}
    K(\bm{x},\bm{x}')=\phi(\bm{x})^\top\cdot \phi(\bm{x}').
\end{equation}
Under this definition, the kernel fulfills all the requirements exposed above. The kernel trick consists then on substituting the product of feature vectors by the kernel function, exploiting the fact that it might be easier to compute the inner product than the vectors themselves. This is especially relevant when the embedding happens in large feature spaces, as it usually happens with quantum kernels. We will use this fact later to define our quantum kernel.

The parameters $\alpha_i$ can be efficiently computed by solving the Lagrangian optimization problem for the soft margin SVM in its dual form: 
\begin{equation}\label{eq:dual_form}
    \arg \max_{\alpha_i}\left(\sum^M_{i=1}\alpha_i-\frac{1}{2}\sum^M_{i=1}\sum^M_{j=1}\alpha_i\alpha_jy_iy_jK(\bm{x}_i,\bm{x}_j)\right),
\end{equation}
subject to $\sum^M_{i=1}\alpha_i y_i=0$ with $0\leq \alpha_i\leq C$ for $i=1,\dots,M$. The penalty term $C>0$ determines the balance between minimizing the training error and maximizing the margin. When $C$ is small, it results in a large margin but potentially high training error.

For this paper, we will make use of the soft margin SVM method from the library Scikit-Learn \cite{pedregosa2011scikit}.  In order to have a competitive benchmark to compare with, we take the classical kernel known as the Radial Basis Function (RBF). This kernel is defined as:
\begin{equation}\label{KerC}
K(\bm{x},\bm{y})=e^{-\gamma||\bm{x}-\bm{y}||^2},
\end{equation}
where $\gamma$ controls the precision of the kernel.

\subsection{Quantum fidelity kernel} \label{sect:QFK}
Once the SVM framework is introduced, we define the quantum kernel that we will consider in this work, based on an analog implementation in an ion trap platform. We start by specifying the dynamical system that will process the input information. 
The Hamiltonian of our system is a Transverse-field Ising model:
\begin{equation}\label{Eq:H}
    H=\sum^N_{i>j=1}\frac{J}{|i-j|^{\alpha}}\sigma_i^x\sigma_j^x+\sum^N_{i=1}h_i\sigma_i^z,
\end{equation}
where $N$ is the number of spins, $h_i$ is the onsite magnetic field for each spin, $\sigma^{a}_i$ ($a=x,y,z$) are the  Pauli matrices and $\alpha$ is a hyperparameter that controls the decay of the spin-spin couplings, with maximum strength $J$ for the nearest-neighbors in the chain. This model can be generated using a combination of an effective magnetic field and Ising interactions in ion trap platforms, with appropriate settings of the spin phases \cite{monroe2021programmable}. Spin-spin interactions are mediated by global laser beams that couple spin and motion according to the Mølmer-Sørensen scheme \cite{molmer1999multiparticle}. The coupling decay is approximated by a power-law, where $\alpha$ can be tuned between $0$ and $3$ by varying parameters of the trap \cite{islam2013emergence,monroe2021programmable}.  Here we will fix $\alpha=1.13$ and $J=1$, following the values adopted by experimentalists to obtain a medium-length interaction range \cite{smith2016many}. 
Our proposal is to use this ion trap to implement a kernel, by encoding the inputs in the local magnetic fields, where site-dependent laser-induced Stark shifts can be used to prepare them \cite{lee2016engineering}, see \cite{smith2016many,zhang2017crystal,morong2021observation} for some examples of experimental implementations. In particular, we introduce the input into the external magnetic field of some or all the spins as $h_i=hx_i$, depending on the dimension of the input vector.  In general:
\begin{equation}\label{Eq:XXZ}
    H(\bm{x})=\sum^N_{i>j=1}\frac{J}{|i-j|^{\alpha}}\sigma_i^x\sigma_j^x+h\sum^N_{i=1}x_i\sigma_i^z,
\end{equation}
where $h$ is the strength of the magnetic field for all spins and $\bm{x}=\{x_1,x_2,...\}$ is the input vector for each input instance with values scaled in the interval $x_i\in[-1,1]$. The chain length $N$ is set to be equal or larger than the number of input features. For less features than the chain length we will set the magnetic field in the remaining spins to zero: for instance, for an input of only two features (and $N\geq 2$),  $\bm{x}=\{x_1,x_2,0,...,0\}$. If $N$ is a multiple of the number of input features, one could redundantly encode the inputs in more than one spin. In this input protocol,  the number of qubits must scale linearly with respect to the number of input features. 

We will now move on to the definition of the quantum kernel.  The system is initialized with all spins in the state $\ket{0}$. Then, we apply the unitary operator of the dynamics:
\begin{equation}
\ket{\psi(\bm{x})}=e^{-iH(\bm{x})\Delta t}\ket{0},
\end{equation}
where $\Delta t$ indicates the time of the simulation. This operation of encoding the input vector into a new Hilbert space is the feature mapping. To construct the kernel, we need a dot product between two vectors in the new feature space. This is expressed as 
\begin{equation}\label{eq:qkernel}
K(\bm{x},\bm{y})=|\braket{\psi(\bm{y})|\psi(\bm{x})}|^2=|\braket{0|e^{iH(\bm{y})\Delta t}e^{-iH(\bm{x})\Delta t}|0}|^2, 
\end{equation}
where we introduced the squared absolute value for experimental purposes. In order to compute the kernel matrix element for inputs $\bm{x}$ and $\bm{y}$, we need to measure the overlap between states $\ket{\psi(\bm{x})}$ and $\ket{\psi(\bm{y})}$. There are several ways of tackling this problem. The most popular one with quantum computers is the quantum kernel estimation method \cite{havlivcek2019supervised}.
This method would require applying the following chain of operations in our system: 
\begin{enumerate}
    \item Initialize the system at $\ket{0}$ through optical pumping.
    \item Apply the sequence of unitary operators $e^{iH(\bm{y})\Delta t}e^{-iH(\bm{x})\Delta t}$, where each unitary is simulated as described after Eq.~\eqref{Eq:H}. Examples of programmed sequences of unitaries in trapped ions can be seen, for instance, in gate-based quantum computing \cite{bruzewicz2019trapped}, Floquet systems \cite{zhang2017crystal} and preparation of thermofield states \cite{zhu2020generation}.
    \item Measure the state of all qubits in the $z$ direction through e.g. fluorescence collected into a CCD camera.
    \item Repeat the whole process to estimate the frequency of obtaining the state $|0\rangle$ in all qubits after measuring, which indeed approximates Eq.~\eqref{eq:qkernel}.
\end{enumerate}  

Let us now describe the different sources of noise that we will study in the system. There are common sources of decoherence for simulations and gate-based computations in trapped ions, such as stray magnetic and electric fields, mode frequency drifts, off-resonant motional excitation, and spontaneous emission \cite{wineland1998experimental}. The main point of decoherence is that it can set a time limit for quantum simulations. Therefore, we model the experimental noise produced by decoherence with a depolarizing channel after the unitary dynamics \cite{hubregtsen2022training,wang2021towards}, which is a simple model that can be easily studied in the context of quantum kernels:
\begin{equation}
    \tilde{\rho}(\bm{x})=(1-p)\rho(\bm{x})+p\frac{I}{2^N},
\end{equation}
where $p$ is the probability of replacing the quantum state by the maximally mixed state $I/2^N$ and $\rho(\bm{x})=\ket{\psi(\bm{x})}\bra{\psi(\bm{x})}$. We assume that $p$ is the same for all inputs $\bm{x}$ to simplify the setting, but this is not necessarily true in general.
 The quantum kernel is given now by the Hilbert-Schmidt product of the feature vectors in the space of density matrices:
\begin{equation}\label{eq:K_tilde}
\tilde{K}(\bm{x},\bm{y})=\text{tr}(\tilde{\rho}(\bm{x})\tilde{\rho}(\bm{y}))=(1-p)^2K(\bm{x},\bm{y})+p(2-p)/2^N.
\end{equation}
Besides, we can add a Gaussian random number to the kernel matrix elements in order to simulate statistical noise due to finite sampling (justified by the central limit theorem):
\begin{equation}
\tilde{K}'(\bm{x},\bm{y})=\tilde{K}(\bm{x},\bm{y})+\xi (\bm{x},\bm{y}),
\end{equation}
where $\xi(\bm{x},\bm{y})$ is generated from a normal distribution of zero mean and standard deviation $s$ for each pair of inputs $(\bm{x},\bm{y})$.  This can be related to the number of measurements $V$ through the relation $s\sim 1/\sqrt{V}$. Since the kernel matrix has to be symmetric, we set $\xi (\bm{x},\bm{y})=\xi (\bm{y},\bm{x})$. In this work, we consider $p=0.01,0.1$ and $s=0.01,0.1$. 

In order to use a kernel matrix for classification tasks, it must be a positive-semidefinite matrix. The two sources of noise we previously introduced might hinder this feature,
so some mitigation techniques should be applied. We use the Tikhonov regularization \cite{hubregtsen2022training}, also known as the shifting technique in this context \cite{wang2021towards}. It consists on shifting all the eigenvalues by the minimum non-positive eigenvalue of $\tilde{K}'$. This is equivalent to subtracting it from the diagonal:
\begin{equation}\label{eq:K_bar}
   \overline{K}=\left\{
                \begin{array}{ll}
                  \tilde{K}'-\lambda_{\text{min}}I \quad &\text{if  } \lambda_{\text{min}}<0\\
                  \tilde{K} '&\text{else}
                \end{array}
              \right.,
\end{equation}
where $\overline{K}$ is now a symmetric positive-semidefinite matrix.
\subsection{Optimization}\label{sect:opt}

As was mentioned in the Introduction, we will introduce an inductive bias in our models by applying bandwidth optimization. To this end, we perform a grid search that will help to visualize the performance for all explored values of hyperparameters. This can be numerically expensive for many hyperparameters (and hard to visualize), so we limit ourselves to grid searches that vary two hyperparameters simultaneously. For the classical case, we vary the scaling factor of the RBF kernel $\gamma$ in Eq.~\eqref{KerC} and the regularization parameter $C$ (introduced after Eq.~\eqref{eq:dual_form}). For the quantum case, we will vary $\Delta t$ and $h$ while fixing $C=1$ (recall we already fixed $J=1$ and $\alpha=1.13$). We will also have a look at the performance as the number of spins increases for $N=2,4,6$.

To measure the performance of our kernel methods, we use the classification accuracy. It is defined as the number of correct predictions (CP) divided by the total number of predictions (TP):
\begin{equation}
    A=\frac{\text{CP}}{\text{TP}}.
\end{equation}
This metric is appropriate since our dataset is well-balanced, with half of each set from each classification class.

In order to find the optimal hyperparameters, we adopted a training, validation, and testing strategy. Training is used to find the optimal parameters of the SVM model for the training set (for a given value of the hyperparameters). The validation set allows us to compute the performance of the model (with the same hyperparameters) for new data that the trained model has not seen. Based on the validation performance, we make the grid search in order to find the optimal hyperparameters. Finally, at the test, we evaluate the final performance of the model, with trained parameters and optimized hyperparameters, for a new set of inputs. 

Three binary classification tasks are evaluated. The Circles (C) and Moons (M) tasks were generated from the library ScikitLearn \cite{pedregosa2011scikit}. These two tasks are paradigmatic examples of benchmark classification tasks because they are easy to generate, while not trivially solvable, and are very popular in the quantum kernel literature \cite{schuld2019quantum,vedaie2020quantum,lloyd2020quantum,li2022quantum,rastunkov2022boosting,yang2023analog}, even with experiments \cite{bartkiewicz2020experimental,kusumoto2021experimental}. The third task, called Ad Hoc (AH), was extracted from the Qiskit dataset library in Python \cite{Qiskit}. This dataset was originally designed to be exactly separable using quantum kernels as proposed in \cite{havlivcek2019supervised}. We evaluate here the capability of the classical RBF and quantum trapped ion kernels to solve these three tasks. 

For the Circles and Moons tasks, Gaussian noise is introduced during the generation of the datasets (with standard deviation equal to 0.2 and 0.3 respectively), in such a way that points belonging to one class can enter into the area that corresponds to the other class (See Figs.~\ref{Fig1} (a) and (b)). For the Ad Hoc task (Fig.~\ref{Fig1} (c)), we set the separation gap as $\Delta=0.3$ (see \cite{havlivcek2019supervised} for more details). Note that the input data noise is the only noise source we considered for the classical kernel.

The datasets are equally divided into three sets (train, validation, and test) with 333 points per set. We remark that given this number of data points and the resolution of the grid search (we used 100 points per hyperparameter), we may find more than one equal-valued minima in the grid search. We need to select one of them to compute the test accuracy. Our choice is to take the middle element of the ordered list of minima, having checked that the other points work equally well.

\section{Numerical results}

\subsection{Classical kernel}

We begin by presenting the results of the classical RBF kernel. 
Table \ref{table:RBF} shows the values of the optimal hyperparameters for the three different tasks, together with the validation accuracy $A_{\text{val}}$ and test accuracy $A_{\text{test}}$. We observe that the values of validation accuracy are larger than those of test accuracy, as expected. The fact that we do not surpass the value $0.9$ of test accuracy in any of the tasks is due to the presence of noise during the generation of the datasets, as explained in the previous section.

\begin{table}[h]
\npdecimalsign{.}
\nprounddigits{2}
\begin{tabular}{ |p{0.8cm}|n{1}{2}|n{1}{2}|n{1}{2}|n{1}{2}|  }
 \hline
 Task & $\gamma$ &$C$ & $A_{\text{val}}$ & $A_{\text{test}}$\\
 \hline
C&2.783e-3&2.984e7&0.910&0.889\\
M&4.977e-2&8.111e6&0.937&0.892\\
AH&2.984e0&2.105e4&0.853&0.841\\
 \hline
\end{tabular}
\npnoround
\caption{Optimal hyperparameters for the classical RBF kernel. The tasks are represented as C (Circles), M (Moons), and AH (Ad Hoc). The values of $A_{\text{val}}$ and $A_{\text{test}}$ represent the accuracy with the optimal hyperparameters for the validation and test sets respectively.}
\label{table:RBF}
\end{table}

\begin{figure}[h]
\captionsetup[subfigure]{}
\begin{center}
\includegraphics[scale=0.95]{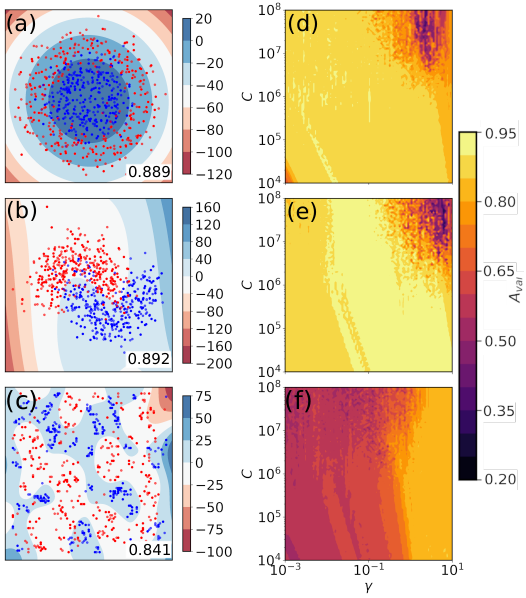}
\caption{Decision function ((a)-(c)) and optimization landscape ((d)-(f)) of the RBF classical kernel model. Figures (a)-(c) were obtained for the test sets by extracting the optimal hyperparameters of figures (d)-(f). The three different tasks are referred to as Circles ((a) and (d)), Moons ((b) and (e)), and Ad Hoc ((c) and (f)). }\label{Fig1}
\end{center}
\end{figure}

Figures \ref{Fig1} (d)-(f) represent the classification accuracy during the validation procedure of the classical kernel over the hyperparameter space. The optimal hyperparameters of Table \ref{table:RBF} were obtained by selecting the values that maximize the validation accuracy, as described in Section \ref{sect:opt}.  For the Circles and Moon tasks (Figs.~\ref{Fig1} (d) and (e)), we observe that a broad combination of hyperparameters allows us to solve them. However, the Ad Hoc task (Fig.~\ref{Fig1} (f)) seems to require a higher nonlinear response from the kernel. Using the kernel with a small scale factor $\gamma$ tends to make the SVM behave like a linear classifier. Conversely, a large scale factor $\gamma$ makes the output classifier highly sensitive to small input changes, which could lead to overfitting even with margin maximization. Figure \ref{Fig1} (f) shows how this task is best solved with the higher values of $\gamma$ (higher nonlinearity) but without overfitting, as the test accuracy of Table \ref{table:RBF} demonstrates.

\subsection{Quantum kernel}

Now we introduce the results of the quantum kernel.
Table \ref{table:XXZ} shows the optimal hyperparameters and classification accuracy of the quantum kernel with the Hamiltonian given in Eq.~\eqref{Eq:XXZ} for the same tasks. This table contains only the noiseless case, i.e., the kernel given by Eq.~\eqref{eq:qkernel}. As in the classical case, we compute the validation and test accuracy for the optimal hyperparameters ($h$ and $\Delta t$), while exploring the number of qubits $N$ as well.   

\begin{table}[h]
\npdecimalsign{.}
\nprounddigits{2}
\begin{tabular}{ |p{0.8cm}|p{0.5cm}|n{2}{2}|n{2}{2}|n{2}{2}|n{2}{2}|  }
 \hline
 \multicolumn{6}{|c|}{No noise} \\
 \hline
 Task & N & $h$ &$\Delta t$ & $A_{\text{val}}$ & $A_{\text{test}}$\\
 \hline
C&2&0.404&2.105&0.763&0.715\\
C&4&0.368&16.298&0.907&0.889\\
C&6&1.874&2.310&0.907&0.874\\
M&2&0.423&49.770&0.637&0.568\\
M&4&2.257&2.783&0.931&0.898\\
M&6&2.984&2.105&0.931&0.898\\
AH&2&1.630&2.205&0.742&0.724\\
AH&4&1.556&6.734&0.835&0.778\\
AH&6&1.556&6.734&0.865&0.823\\
 \hline
\end{tabular}
\caption{Optimal hyperparameters for the quantum kernel model without noise. The tasks are represented as C (Circles), M (Moons), and AH (Ad Hoc). $N$ represents the number of qubits, $h$ is the strength of the external magnetic field and $\Delta t$ is the unitary evolution's time scale. The values of $A_{\text{val}}$ and $A_{\text{test}}$ represent the accuracy with the optimal hyperparameters for the validation and test sets respectively.}
\label{table:XXZ}
\end{table}

 \begin{table*}[ht!]
\npdecimalsign{.}
\nprounddigits{2}

\begin{tabular}{ |p{0.8cm}|p{0.5cm}|n{2}{2}|n{2}{2}|n{2}{2}|n{2}{2}|n{2}{2}|n{2}{2}|n{2}{2}|n{2}{2}|n{2}{2}|n{2}{2}|n{2}{2}|n{2}{2}|n{2}{2}|n{2}{2}|n{2}{2}|n{2}{2}|}
 \hline
\multicolumn{2}{|c}{} & \multicolumn{4}{c}{$s=0.01$ and $p=0.01$} & \multicolumn{4}{c}{$s=0.1$ and $p=0.01$} & \multicolumn{4}{c}{$s=0.01$ and $p=0.1$} & \multicolumn{4}{c|}{$s=0.1$ with $p=0.1$} \\
 \hline
 Task & N & $h$ &$\Delta t$ & $A_{\text{val}}$ & $A_{\text{test}}$& $h$ &$\Delta t$ & $A_{\text{val}}$ & $A_{\text{test}}$& $h$ &$\Delta t$ & $A_{\text{val}}$ & $A_{\text{test}}$& $h$ &$\Delta t$ & $A_{\text{val}}$ & $A_{\text{test}}$\\
 \hline
C&2&0.183&19.630&0.781&0.721&0.705&7.391&0.763&0.706&0.486&1.918&0.775&0.706&0.705&7.391&0.757&0.706  \\
C&4&1.024&2.535&0.907&0.883&0.278&29.836&0.898&0.883&0.890&2.915&0.907&{\boldmath}0.889&0.423&25.950&0.904&0.883\\
C&6&0.559&6.734&0.907&{\boldmath}0.886&0.739&4.642&0.901&0.883&0.774&5.094&0.907&0.880&0.368&59.948&0.901&{\boldmath}0.892\\
M&2&1.417&2.310&0.637&0.583&1.874&2.535&0.616&0.574&1.556&2.105&0.628&0.598&1.292&3.199&0.628&0.544\\
M&4&1.177&8.902&0.931&0.874&1.789&3.054&0.910&{\boldmath}0.898&1.963&2.915&0.931&{\boldmath}0.898&2.257&2.105&0.913&0.892\\
M&6&2.848&2.205&0.931&{\boldmath}0.895&1.874&3.352&0.922&{\boldmath}0.901&2.848&2.205&0.931&{\boldmath}0.895&1.707&3.199&0.910&{\boldmath}0.898\\
AH&2&1.789&2.009&0.739&0.703&3.765&1.000&0.676&0.676&1.789&2.009&0.736&0.703&1.789&2.009&0.682&0.703\\
AH&4&1.417&7.055&0.823&0.778&2.364&8.498&0.724&0.763&1.556&6.734&0.814&0.778&1.417&7.391&0.715&0.799\\
AH&6&1.292&8.498&0.868&0.820&1.177&10.235&0.757&0.802&1.556&18.738&0.859&{\boldmath}0.859&1.417&7.391&0.751&0.820\\
\hline
\end{tabular}
\caption{Optimal hyperparameters for the quantum kernel model with noise. The tasks are represented as C (Circles), M (Moons) and AH (Ad Hoc). $N$ represents the number of qubits, $h$ is the strength of the external magnetic field and $\Delta t$ is the unitary evolution's time scale. The values of $A_{\text{val}}$ and $A_{\text{test}}$ represent the accuracy with the optimal hyperparameters for the validation and test sets respectively.}
\label{table:XXZnoise}
\end{table*}

The first thing that calls our attention from the results presented in Table \ref{table:XXZ} is that the performance significantly improves when increasing the number of qubits from $N=2$ to $N=4$ in all tasks. This transition is directly related to the size of the quantum feature space where the inputs are introduced, increasing the expressivity of the model. However, the Ad Hoc task is the only one that also displays a significant change when increasing the number of spins from $N=4$ to $N=6$. This suggests that the task at hand determines the optimal size of the Hilbert space, depending on the expressiveness required. 

Regarding the effect of statistical (s) and depolarization (p) noise, we analyze four different situations where the quantum kernel is given by Eqs.~\eqref{eq:K_tilde}-\eqref{eq:K_bar}, summarized in Table \ref{table:XXZnoise}. In the first case, we added a small amount of both statistical and depolarizing noise, with $s=0.01$ and $p=0.01$. Here the test accuracy is slightly decreased in almost all the cases, but we also observe that it can be also slightly increased, as in the Circles task with $N=6$.  Then, we increase the statistical noise to $s=0.1$, finding a small negative effect over validation accuracy that was not present with $s=0.01$. Instead, when increasing the depolarizing noise to $p=0.1$ while keeping $s=0.01$, both validation and test accuracy return to values similar to the noiseless case, with even higher performance in the Ad Hoc task for $N=6$. Finally, the case of larger statistical and depolarizing noise ($s=0.1$ and $p=0.1$) shows that test performance is almost identical to the noiseless situation, even larger in some cases. However, the validation accuracy decreases again as in the case of $s=0.1$ and $p=0.01$, meaning that this effect is produced by the statistical noise. 

\begin{figure}[h]
\captionsetup[subfigure]{}
\begin{center}
\includegraphics[scale=0.95]{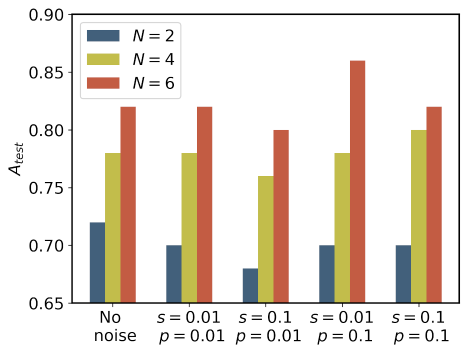}
\caption{Test accuracy of the quantum kernel model for the Ad Hoc task. The color of each bar corresponds to a given number of qubits and each group of bars corresponds to a set of noise parameters.}\label{FigADHOC}
\end{center}
\end{figure}

To visualize the change in performance with noise, Fig.~\ref{FigADHOC} represents the performance of the quantum kernel for the Ad Hoc task in all the studied situations. The $x$ axis describes the noise parameters and the $y$ axis represents the test accuracy. The color of each bar corresponds to a given number of qubits. Figure \ref{FigADHOC} illustrates our main findings, namely that test accuracy increases with system size and that statistical noise can be detrimental, although an appropriate amount of depolarizing noise can compensate for this negative effect, even finding the best performance for the Ad Hoc task for $N=6$ with $s=0.01$ and $p=0.1$. 

When comparing the optimal quantum kernels with the classical kernel, the quantum models for $N=4$ and $N=6$ obtain a very similar performance to the classical case, even under the effect of noise. For the Ad Hoc task, we can also find one situation where the test accuracy of the quantum kernel is larger than in the classical case ($N=6$ for $s=0.01$ and $p=0.1$). On the one hand, the fact that statistical noise (less measurement precision) tends to decrease the performance is an evident negative effect that can have severe consequences, such as the loss of any possible generalization advantage of quantum kernels \cite{wang2021towards}. However, as shown in \cite{liu2021rigorous}, quantum kernels with a controlled additive sampling noise in SVM are robust, where the noisy hyperplane is close to the noiseless hyperplane. On the other hand, depolarizing noise corresponds to a loss of information about the quantum state with probability $p$, so large values of $p$ would also hinder the performance of quantum devices. But as we observe here, it can act as a regularization parameter for small values. We remark that in any case, techniques can be carried out to mitigate the effect of both sources of noise \cite{wang2021towards,hubregtsen2022training,moradi2022error}.

\begin{figure}[htb!]
\captionsetup[subfigure]{}
\begin{center}
\includegraphics[scale=0.95]{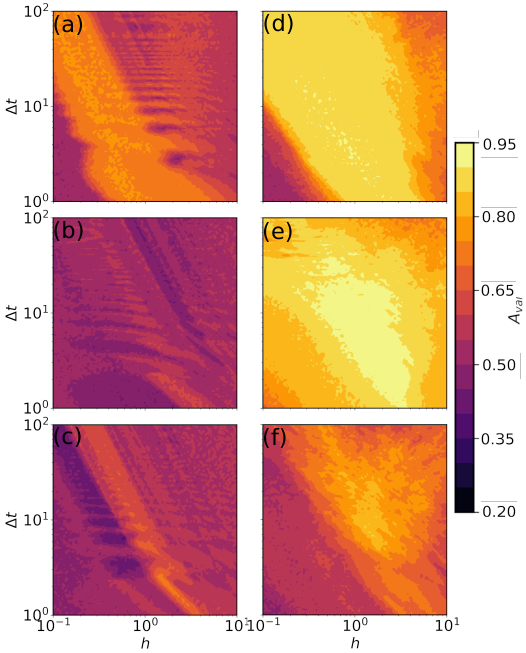}
\caption{Optimization landscape for the quantum kernel model with the validation datasets. The statistical noise strength is given by $s=0.01$ and the depolarizing noise parameter is $p=0.01$. The left column corresponds to $N=2$ while the right column corresponds to $N=6$. The top row is the Circles task, the middle row is the Moons task, and the bottom row is the Ad Hoc task.}\label{Fig4}
\end{center}
\end{figure}

We now proceed to visualize the relationship between the optimal hyperparameters and task performance. Figure \ref{Fig4} shows the validation accuracy in terms of the hyperparameters $h$ and $\Delta t$ for the systems associated with $s=0.01$ and $p=0.01$ in Table \ref{table:XXZnoise}. We notice that the optimization landscape is qualitatively similar in all the studied situations, so Fig.~\ref{Fig4} is a representative example. The left and right columns correspond to $N=2$ and $N=6$ respectively, while each row corresponds to a different task. As already observed in Table \ref{table:XXZnoise}, the validation accuracy reaches higher values for $N=6$ than for $N=2$ due to the larger Hilbert space. But Fig.~\ref{Fig4} also shows that the region of hyperparameters that could solve the tasks may be broad. 
The Circles task (Fig.~\ref{Fig4} (d)) exhibits a wide range of hyperparameters yielding high accuracy, with a possible linear trend for $h \Delta t\sim \text{cte}$ in the center. The Moons task (Fig.~\ref{Fig4} (e)) reveals an even wider range of successful hyperparameters. Interestingly, also for the Ad Hoc task (Fig.~\ref{Fig4} (f)), optimal hyperparameters consistently cluster in the center of the figures. Our analysis illustrates that these central hyperparameter combinations lead to high kernel expressivity.

One may try to explain the previous results by establishing a connection between the value of the optimal hyperparameters and the underlying dynamical features of the model. However, it seems more complicated than that. Notice that the kernel of Eq.~\eqref{eq:qkernel} has three hyperparameters related to the dynamics, $J$, $h$, and $\Delta t$, but in fact, only two of them are independent (when combined as $J\Delta t$ and $h\Delta t$ in the unitary operator). Then the coefficient $J/h$ indicates the relative weight of each of the two terms of the Hamiltonian in Eq.~\ref{Eq:XXZ}. Recalling that we previously fixed $J=1$, our results suggest that the relative weight of the external magnetic field term to the coupling term would only become a relevant factor in solving some specific tasks. For instance, the Moons task does not strongly depend on the rate $J/h$: in Fig.~\ref{Fig4} (e) we observe that values of the hyperparameters such as $0.3<h<3$ with $\Delta t\sim 10$ provide a similar performance. That is, the rates $1/3<J/h<3$ work equally well. On the contrary, the Ad Hoc task requires a very specific ratio $J/h$: Fig.~\ref{Fig4} (f) shows a clear optimal region of hyperparameters in the center of the figure, around $\Delta t\sim 10$ and $h\sim 1$.

\section{Discussion}

 Kernel methods are a promising direction in the search for applications of quantum machine learning techniques. In particular, classically intractable quantum simulation models can be exploited to explore quantum advantages. Our work proposes ion trap platforms as an analog experimental platform to compute quantum kernels, demonstrating their viability through numerical simulations of realistic scenarios with depolarizing and statistical noise. 

The computation of quantum kernels proposed here for trapped ions builds on the kernel estimation protocol of Ref. \cite{havlivcek2019supervised}.
Computing the unitary dynamics given by $-H(\bm{y})$ only requires the change of sign of the couplings and the definition of new magnetic fields depending on the input $\bm{y}$. One potential challenge is whether it is possible to make this sudden change between $H(\bm{x})$ and $-H(\bm{y})$ on the fly. There are other possible routes to measure the overlap between two quantum states that might be applied to this quantum simulation platform. For example, the SWAP test and extensions \cite{buhrman2001quantum,fanizza2020beyond} would allow evolving in parallel the states $\ket{\psi(\bm{x})}$ and $\ket{\psi(\bm{y})}$  without the sudden quench. Classical machine learning techniques can help here \cite{cincio2018learning,cincio2021machine}, and full tomography allows us to obtain the quantum kernel as well \cite{heyraud2022noisy}. Finally, a promising direction might be to use randomized measurements \cite{elben2020cross,haug2021quantum}, where the overlap of states generated by independent Hamiltonians can be measured.

 We evaluated the performance of the proposed quantum kernel in three different binary classification tasks and compared the results with the well-known RBF classical kernel. We made a grid search of optimal hyperparameters for both classical and quantum models, also exploring the number of qubits in the latter. The bandwidth optimization of the quantum model introduces an inductive bias that restricts the type of functions that can be solved. We find that both classical and quantum models reach a very similar performance, closely saturating the maximum possible test accuracy for the given test sets without overfitting. This means that the proposed quantum model is operational for classification tasks, being robust under the presence of different sources of noise. 

The search for optimal quantum kernels was carried out for only two hyperparameters with a very specific Hamiltonian (eq.~\eqref{Eq:H}) and input scheme, but more possibilities can be explored. The hyperparameters $\alpha$ and $C$ can be also tuned, the input can be redundantly injected (like $\bm{x}=\{x_1,x_2,x_1,x_2,...\}$ in eq.~\eqref{Eq:H}), or we can even add new terms to the Hamiltonian such as a magnetic field in the $x$ axis.  We tested these possibilities for the presented tasks, reaching a similar or worse performance. However, we do not discard that a more systematic evaluation of these strategies could bring an improvement. In fact, the quantitative performance in different tasks is expected to vary under different interactions or driving in the Hamiltonian.
 
 Future work will be devoted to exploring the effect of increasing the number of qubits to solve tasks with a larger number of input features. On the one hand, one can expand the number of features that can be codified in a single qubit by feeding them over time with an input-dependent driving field, as presented in Ref.~\cite{heyraud2022noisy}. On the other hand, projected quantum kernels can be explored in this setting \cite{huang2021power}, limiting the effective Hilbert space dimension to maximize the generalization ability of the model.

\section*{ACKNOWLEDGMENTS}
 We acknowledge the Spanish State Research Agency, through the Mar\'ia de Maeztu project CEX2021-001164-M funded by the MCIN/AEI/10.13039/501100011033 and through the COQUSY project PID2022-140506NB-C21 and -C22 funded by MCIN/AEI/10.13039/501100011033, MINECO, through the QUANTUM SPAIN project, and EU through the RTRP - NextGenerationEU within the framework of the Digital Spain 2025 Agenda. The CSIC Interdisciplinary Thematic Platform (PTI) on Quantum Technologies in Spain is also acknowledged. RMP acknowledges the QCDI project funded by the Spanish Government and part of this work was also funded by MICINN/AEI/FEDER and the University of the Balearic Islands through a pre-doctoral fellowship (Grant No. MDM-2017-0711-18-1).

\end{document}